\documentclass[prl,aps,amsmath,amssymb,superscriptaddress,twocolumn]{revtex4-1}
\usepackage{graphicx}
\usepackage{dcolumn}
\usepackage{bm}
\usepackage{xcolor}
\usepackage[latin9]{inputenc}
\usepackage{amsmath}
\usepackage{multirow}
\usepackage[per-mode=symbol]{siunitx}
\DeclareSIUnit\gauss{G}
\graphicspath{{/}}

\newcommand{\ket}[2][]{\mathinner{\lvert#2\rangle}_{\hspace{-0.1em}#1}}

\newcommand{\ayop}{\hat{a}_y}
\newcommand{\adyop}{\hat{a}^\dagger_y}

\newcommand{\aop}{\hat{a}}
\newcommand{\adop}{\hat{a}^\dagger}
\newcommand{\Fp}{\hat{F}_{+}}
\newcommand{\Fm}{\hat{F}_{-}}

\newcommand{\Fy}{\hat{F}_{y}}

\begin{document}


\title{Observation of ultra-strong spin-motion coupling for cold atoms in optical microtraps}

\author{A.~Dareau}
\author{Y.~Meng}
\author{P.~Schneeweiss}
\email{philipp.schneeweiss@tuwien.ac.at}
\affiliation{%
 Vienna Center for Quantum Science and Technology,\\
 TU Wien -- Atominstitut, Stadionallee 2, 1020 Vienna, Austria
}%
\author{A.~Rauschenbeutel}
\email{arno.rauschenbeutel@hu-berlin.de}
\affiliation{%
 Vienna Center for Quantum Science and Technology,\\
 TU Wien -- Atominstitut, Stadionallee 2, 1020 Vienna, Austria
}%
\affiliation{%
 Department of Physics, Humboldt-Universit\"at zu Berlin, 10099 Berlin, Germany
}%

\date{\today}

\begin{abstract}
We realize a mechanical analogue of the Dicke model, achieved by coupling the spin of individual neutral atoms to their quantized motion in an optical trapping potential. The atomic spin states play the role of the electronic states of the atomic ensemble considered in the Dicke model, and the in-trap motional states of the atoms correspond to the states of the electromagnetic field mode. The coupling between spin and motion is induced by an inherent polarization gradient of the trapping light fields, which leads to a spatially varying vector light shift. We experimentally show that our system reaches the ultra-strong coupling regime, i.e., we obtain a coupling strength which is a significant fraction of the trap frequency. Moreover, with the help of an additional light field, we demonstrate the in-situ tuning of the coupling strength. Beyond its fundamental interest, the demonstrated one-to-one mapping between the physics of optically trapped cold atoms and the Dicke model paves the way for implementing protocols and applications that exploit extreme coupling strengths.
\end{abstract}

\maketitle

The quantum Rabi model (QRM) describes the interaction of a two-level emitter with a single quantized mode of the electromagnetic field or, more generally, of a two-level system (TLS) with a bosonic mode. Together with its extension for an ensemble of emitters, i.e., the Dicke model (DM), it constitutes a cornerstone of quantum optics~\cite{Braak16}. The physics predicted by the QRM and the DM strongly depends on the relative values of the mode frequency, $\omega$, and the coupling strength between the TLS and the bosonic mode, $g$. For weak coupling, i.e., $g/\omega \ll 1$, the rotating wave approximation (RWA) applies. In this case, the QRM and the DM reduce to the Jaynes-Cummings and the Tavis-Cummings models, respectively. The RWA breaks down in the ultra-strong coupling regime (USC), i.e., for $g/\omega \gtrsim 0.1$. When increasing the coupling strength further, one enters the deep-strong coupling regime (DSC)~\cite{Rossatto17}. For such high values of $g/\omega$, new phenomena are expected~\cite{Ashhab10, Casanova10, Garziano16, ArXiv_forn-diaz18, ArXiv_Kockum18}. The existence of a quantum phase transition in the thermodynamic limit adds to the richness of the DM~\cite{Hepp73, Wang73, Emary03b}. Furthermore, USC and DSC may enable novel protocols for quantum communication and quantum information processing~\cite{Nataf11,Romero12,Kyaw15}.

Over the last decade, USC was reached using various experimental platforms~\cite{Anappara09,Guenter09,Bourassa09,Todorov10,Niemczyk10,Forn-Diaz10,Schwartz11,Zhang16,George16}. More recently, DSC was achieved in circuit quantum electrodynamics~\cite{Yoshihara17,Forn-Diaz17} as well as by coupling a THz metamaterial with cyclotron resonances in a two-dimensional electron gas~\cite{Bayer17}. While these systems reach record-high ratios of $g/\omega$, the large coupling strengths make state preparation and read-out challenging. For this reason, alternative routes were proposed to achieve large coupling in experimental platforms that, at the same time, offer a high level of control and tunability. Following this path, the dynamics of the QRM in the USC and DSC regimes was studied, respectively, with analog and digital quantum simulations using circuit quantum electrodynamics~\cite{Braumueller17,Langford17}, and DSC was reached with single trapped ions~\cite{Lv18}. 

Here, we implement a mechanical analogue of the Dicke model by coupling the spin of individual neutral atoms to their quantized motion in a trapping potential. In our approach, the coupling is enabled by spatial gradients of the vector light shift inherent to optical microtraps. Fluorescence spectroscopy grants access to the energy spectrum of the system, revealing ultra-strong spin-motion coupling in our experiment, i.e., the coupling strength is a significant fraction of the mode frequency. Furthermore, we demonstrate that the coupling strength can be readily and independently tuned in situ.

\begin{figure}
	\includegraphics[width=0.9\columnwidth]{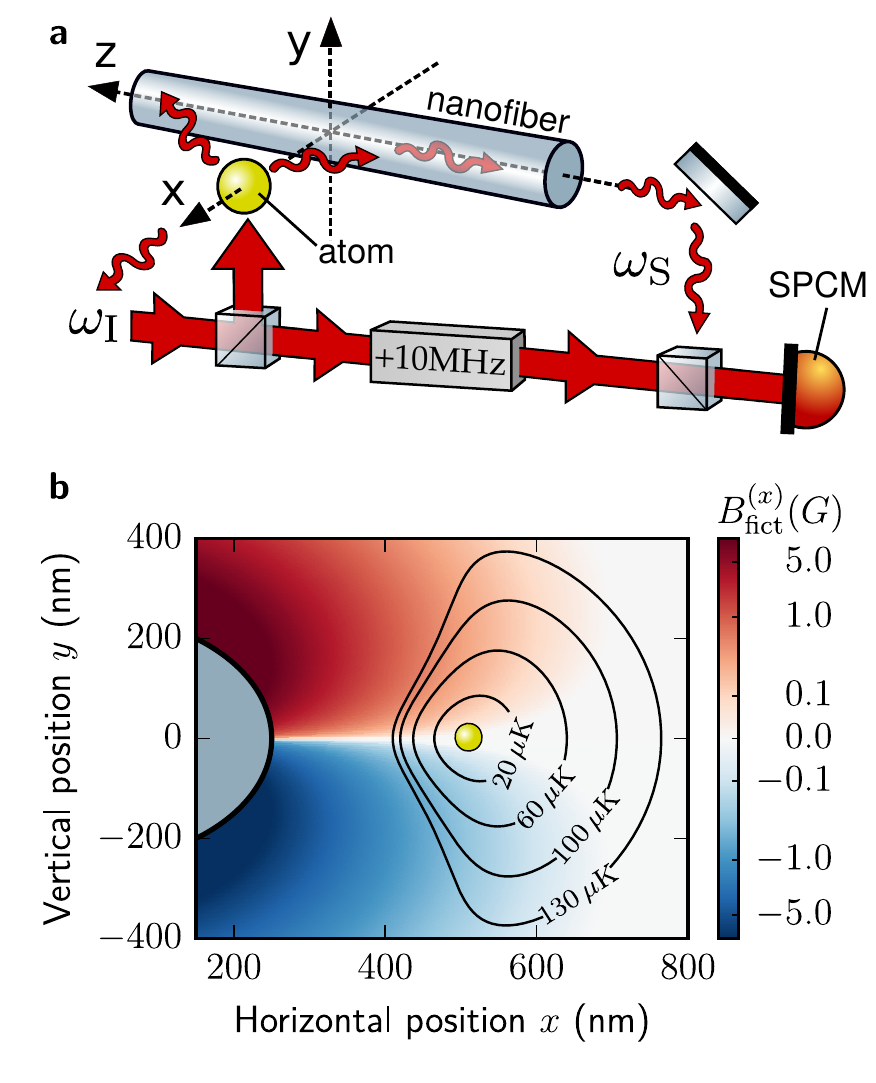}
	\caption{\textbf{Experimental setup}. \textbf{a}, Individual cesium atoms are trapped near the surface of the nanofiber-section of a tapered optical fiber. They are exposed to a near-resonant excitation laser field (frequency $\omega_\mathrm{I}$), propagating along the $+y$-direction. A fraction of the atomic fluorescence is scattered into the guided mode of the nanofiber (frequency $\omega_\mathrm{S}$). This light is superposed with a reference beam that is derived from the excitation laser and frequency-shifted by $+10\,\mathrm{MHz}$. The resulting beat note is recorded with a single photon counting module (SPCM). Its Fourier analysis yields the fluorescence spectrum, which grants access to the energy spectrum of the trapped atoms. \textbf{b}, The strong spatial confinement of the trapping light fields results in large fictitious magnetic field gradients that give rise to a coupling between the atomic spin and its motional degree of freedom. Contours of the scalar part of the trapping potential are indicated by black lines. A yellow dot marks the position of the atom at one trapping site. The amplitude of the main component of the field, $B_\mathrm{fict}^{(x)}$, is shown as a density plot (logarithmic color scale). The gray disk indicates the cross section of the nanofiber.}
	\label{fig:ExpSetup_combined}
\end{figure}

Our implementation employs laser-cooled individual cesium atoms trapped in the evanescent light field surrounding the nanofiber-section of a tapered optical fiber, see Fig.~\ref{fig:ExpSetup_combined}a and supplemental material (SM). The strong transverse confinement of the trapping light fields results in a strong polarization gradient in the azimuthal direction. In addition to the scalar light shift that gives rise to trapping, atoms in the evanescent field then experience a spatially-varying vector light shift~\cite{Albrecht16, Meng18}. This shift can be thought of as arising from the Zeeman interaction with a position-dependent fictitious magnetic field, $\bm{B}_\mathrm{fict}$~\cite{Cohen-Tannoudji72}. For our configuration, $\bm{B}_\mathrm{fict}$ mainly points along the $x$-direction, and its amplitude exhibits strong spatial gradients, see Fig.~\ref{fig:ExpSetup_combined}b. Near the trap minimum, the $x$-component of the fictitious magnetic field varies approximately linearly along $y$, so that $\bm{B}_\mathrm{fict} \approx b_y y \,\bm{e}_x$, with $\bm{e}_x$ the unit vector along $x$ and $b_y \approx \SI{1.9}{\gauss\per\micro\meter}$.

The Zeeman interaction of a trapped atom with this fictitious magnetic field results in a coupling between the atomic spin and motional degrees of freedom (DOF). Here, we assume a harmonic trapping potential, with a set of frequencies $\{\omega_i\}$ and annihilation operators $\{\aop_i\}$ ($i=x,y,z$). In addition to $\bm{B}_\mathrm{fict}$, we apply a homogeneous offset magnetic field, $\bm{B}_0 = B_0 \, \bm{e}_{y}$, along the $y$-direction. The dynamics of a trapped atom is then described by the following Hamiltonian:

\begin{equation}
 \hat{H} = \sum\limits_{i=x,y,z} \hbar \omega_i \adop_i \aop_i + g_F \mu_B \bm{\hat{F}}\cdot(\bm{B}_0+\bm{B}_{\text{fict}}),
\label{eq:Hat_ini}
\end{equation}

\noindent with $g_F$ the hyperfine Land\'e factor and $\mu_B$ the Bohr magneton. Assuming that the fictitious magnetic field consists of a linear gradient along $y$, and only considering the $y$ motional DOF, we can rewrite~\eqref{eq:Hat_ini} as (see SM):

\begin{equation}
 \hat{H}_y = \hbar \omega_y \adyop \ayop + \hbar\Delta\Fy + \frac{\hbar g_y}{\sqrt{2F}} \left( \adyop + \ayop \right) \left( \Fp + \Fm \right),
\label{eq:Hat_QRM}
\end{equation}

\noindent where $\Fp$ ($\Fm$) is the spin raising (lowering) operator for the eigenstates of $\Fy$ with eigenvalues $\hbar m_{F}$. For $F=1/2$, Hamiltonian~\eqref{eq:Hat_QRM} corresponds to the QRM, while for $F>1/2$, as is the case for cesium, it corresponds to the DM. The physics is governed by three parameters: The bosonic mode frequency, $\omega_y$, the Zeeman splitting between adjacent $m_{F}$-states, $\Delta \propto B_0$, and the spin-motion coupling strength, $g_y \propto b_y$. For our configuration, we expect $g_y \approx 2\pi\times\SI{19}{\kilo\hertz}$ for a calculated trap frequency $\omega_y \approx 2\pi\times\SI{95}{\kilo\hertz}$, i.e., $g_y/\omega_y \approx 0.2$.

\begin{figure*}
	\includegraphics[width=0.9\textwidth]{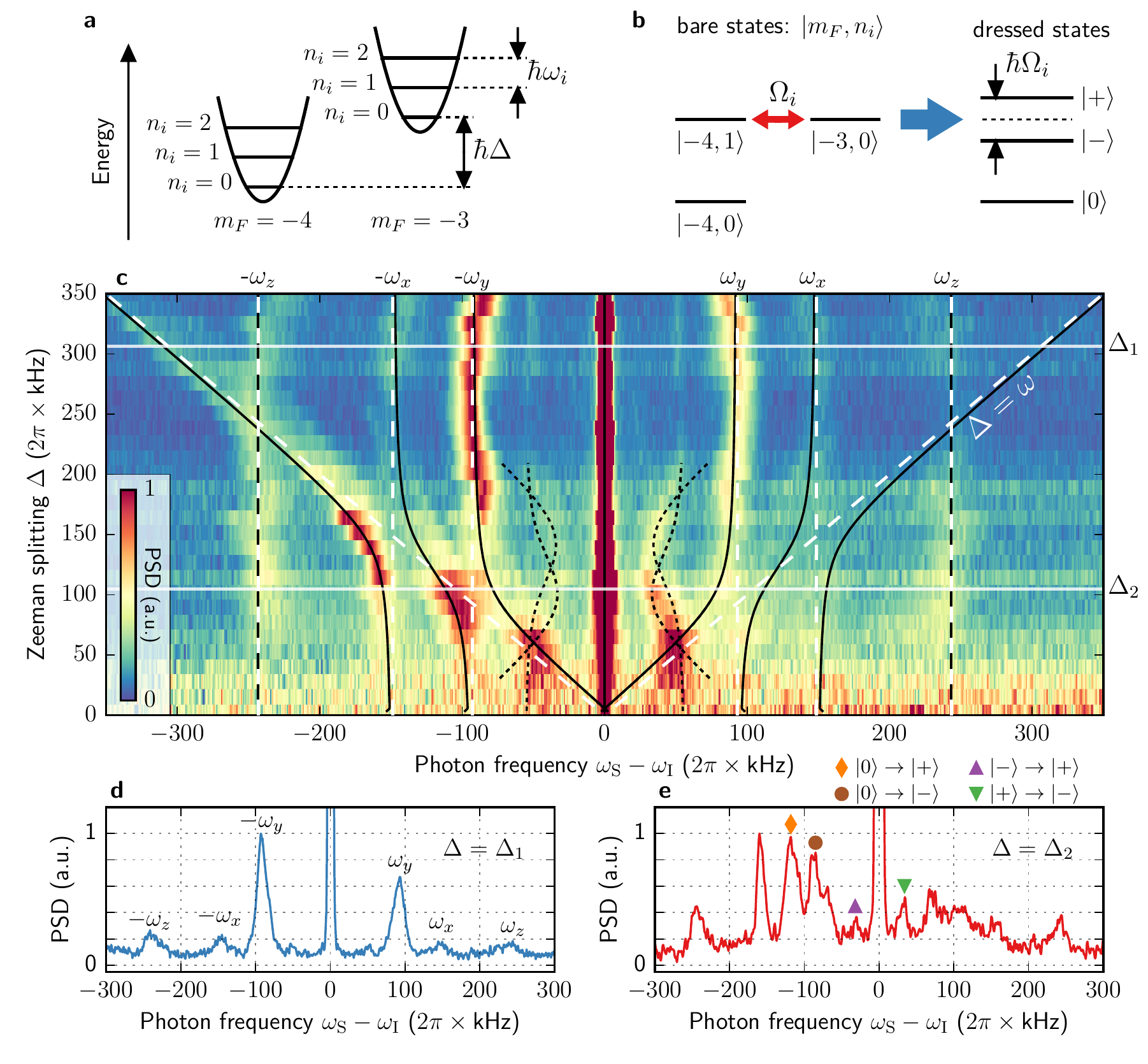}
	\caption{\textbf{Experimental signature of ultra-strong spin-motion coupling}. \textbf{a}, The bare eigenstates in the harmonic trapping potential are $\ket{m_F, n_i}$ with eigenenergies $\hbar(m_F \Delta+n_i \omega_i)$, where  $\hbar\Delta$ is the Zeeman splitting between two neighboring $m_F$-states and $\hbar\omega_i$ is the energy of one motional quantum. The spin-motion coupling is resonant for $\Delta = \omega_i$. \textbf{b}, At resonance, the spin-motion coupling lifts the degeneracy between the bare states $\ket{-4,1}$ and $\ket{-3,0}$, and the new eigenstates $\ket{+}$ and $\ket{-}$ (the dressed states) are split by $\hbar \Omega_i$. \textbf{c}, Fluorescence spectra for different values of $\Delta$. Avoided crossings occur when the resonance condition is fulfilled for the $x$- and $y$-DOFs. Dashed white lines indicate the frequencies of the transitions between bare states and are derived from a fit of the experimental data far from resonance (see SM). Black lines correspond to transitions between dressed states, where the coupling strengths are derived from a fit of the experimental data at resonance (see SM). The values of the fit parameters are given in the main text. Solid black lines correspond to transitions from and to the ground state and dashed black lines to transitions between excited states. \textbf{d}, \textbf{e}, Fluorescence spectra, measured for two Zeeman splittings, $\Delta_1$ and $\Delta_2$, respectively (cf.~solid horizontal white lines in \textbf{c}). Far from resonance, three pairs of motional sidebands are apparent (\textbf{d}). When the coupling is resonant, one motional sideband is split (\textbf{e}). We also observe sidebands corresponding to transitions between the excited states (green and purple triangles).}
	\label{fig:2D_Spectrum}
\end{figure*}

The low-energy eigenstates of $\hat{H}_y$ are illustrated in Fig.~\ref{fig:2D_Spectrum}a,b. We consider the case of cesium in the $F=4$ hyperfine ground state. In the absence of spin-motion coupling ($g_y=0$), the eigenstates are the bare states $\ket{m_F, n_y}$, where  $n_y$ corresponds to a Fock state of the harmonic trapping potential. In the presence of spin-motion coupling, the new eigenstates are dressed states. When the coupling is resonant ($\Delta = \omega_y$), the degeneracy of the bare states $\ket{-4,1}$ and $\ket{-3,0}$ is lifted, and the new eigenstates are $\ket{\pm}=(\ket{-4,1} \mp \ket{-3,0})/\sqrt{2}$, separated  in energy by $\hbar\Omega_y$, where $\Omega_y>0$ is the Rabi frequency. Here, we expect $\Omega_y = 2 g_y\approx 2\pi\times\SI{38}{\kilo\hertz}$ (see SM). 

In order to probe the low-energy part of Hamiltonian~\eqref{eq:Hat_QRM}, we perform a heterodyne fluorescence spectroscopy measurement~\cite{Jessen92, Meng18}. The experimental setup is sketched in Fig.~\ref{fig:ExpSetup_combined}a. The atoms are exposed to a laser light field propagating along the $+y$-axis and $\sigma^-$-polarized with respect to the propagation direction. The laser is red-detuned with respect to the cycling transition of the $D_2$ line of Cesium, and its intensity is kept low enough to ensure that it is scattered coherently by the atoms (see SM). This laser provides degenerate Raman cooling~\cite{Meng18} and optical pumping, so that most of the atoms populate the low-lying energy states depicted in Fig.~\ref{fig:2D_Spectrum}a,b. Part of the fluorescence light is scattered into the guided mode of the optical nanofiber~\cite{Mitsch14b}. This light is superposed with a reference beam, derived from the excitation laser and frequency-shifted by $+\SI{10}{\mega\hertz}$. The resulting beat note is recorded using a single photon counting module (SPCM). Post-processing the SPCM data yields the intensity power spectral density (PSD). This heterodyne setup enables a precise measurement of the frequency difference between the incoming photons from the excitation beam (frequency $\omega_I$) and the photons scattered by the atoms (frequency $\omega_S$). In the case of elastic scattering, the atomic state and the frequency of the photons are unchanged ($\omega_I = \omega_S$), yielding the carrier peak in the PSD. In the case of inelastic scattering, the atomic state is changed and the difference of energy between the incoming and scattered photons has to match the difference of energy between the initial and final atomic states. This gives rise to sidebands around the carrier peak, the positions of which grant access to the energy spectrum of the atoms.

We record fluorescence spectra for different values of the Zeeman splitting, $\Delta\propto B_0$, see Fig.~\ref{fig:2D_Spectrum}c-e. Far from resonance, i.e., for $|\Delta - \omega_i| \gg \Omega_i$, transitions between the bare states result in three pairs of motional sidebands, see Fig.~\ref{fig:2D_Spectrum}d. These transitions change the motional state of the atom but not its spin. These sidebands do not depend on $\Delta$, and their positions can be used to infer the trap frequencies. We find $\left\{\omega_x,\, \omega_y,\, \omega_z\right\} = 2\pi\times \left\{149(2),\, 93(2),\, 243(5)\right\}\,\mathrm{kHz}$. The strong asymmetry of the amplitudes of the positive- and negative-frequency peaks indicates that the atoms are close to the motional ground state~\cite{Meng18}. A fourth peak is also visible in the upper left part of Fig.~\ref{fig:2D_Spectrum}c. It corresponds to a transition between adjacent $m_F$-states for a given motional state. Its position depends linearly on $\Delta$. Close to resonance, we observe a splitting of the motional sideband corresponding to the resonantly coupled DOF. This is clearly visible in Fig.~\ref{fig:2D_Spectrum}e, which is measured close to the resonance of the $y$-DOF, i.e., for $\Delta \approx \omega_y$. The width of the splitting already indicates that we operate in the USC regime. When scanning $\Delta$ around resonance, an avoided crossing is observed. Such an avoided crossing is also visible for the $x$-DOF, indicating that strong spin-motion coupling is present for this DOF, too. This additional coupling could arise from the polarizations of the trapping light fields not being perfectly aligned and/or from a spurious vector light shift originating from the interference of the probe light with its reflection on the nanofiber. Both effects should lead to a significant $x$-gradient of the fictitious magnetic field and, thus, to a strong spin-motion coupling for this DOF. 

\begin{figure}
	\includegraphics[width=0.9\columnwidth]{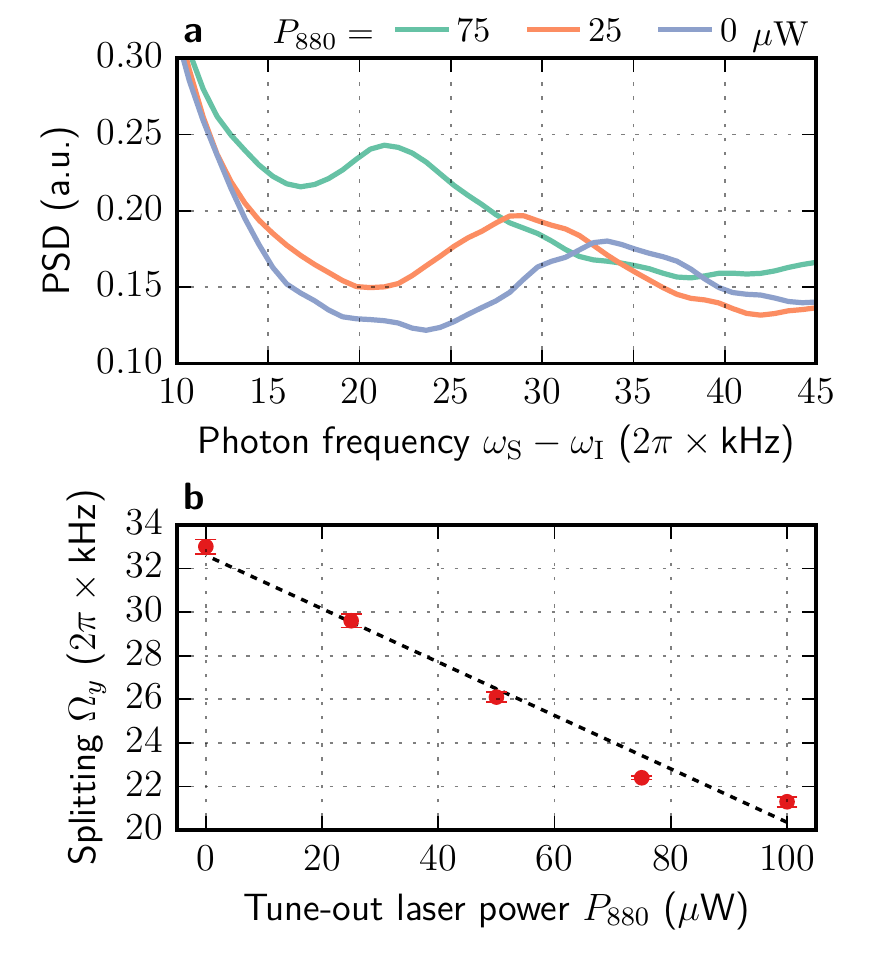}
	\caption{\textbf{Experimental demonstration of the tunability of the spin-motion coupling strength}. A nanofiber-guided tune-out laser field at a wavelength of $\lambda=\SI{880}{nm}$ allows one to modify the fictitious magnetic field gradient along $y$ and, thus, the corresponding spin-motion coupling strength, without changing the scalar part of the trapping potential. \textbf{a}, Fluorescence spectra, taken for resonant coupling of the $y$-DOF ($\Delta = \omega_y$) and different values of the tune-out laser power, $P_{880}$. We show the positive-frequency peak corresponding to the transition between the dressed states. Its position corresponds to the Rabi splitting, $\Omega_y$. This peak shifts towards the carrier for increasing values of $P_{880}$, indicating a reduction of the coupling strength. \textbf{b}, As expected from theory, $\Omega_y$ depends linearly on $P_{880}$, see red circles. A fit (black dashed line) yields $\mathrm{d}\Omega_y/\mathrm{d}P_{880} = \SI{-120(10)}{\hertz\per\micro\watt}$. For $P_{880} > \SI{100}{\micro\watt}$, the proximity of the carrier impedes a precise measurement of the peak position.}
	\label{fig:880_effect}
\end{figure}

Besides the Rabi splitting, a new pair of peaks is apparent close to resonance. These sidebands, labeled by triangles in Fig.~\ref{fig:2D_Spectrum}e, are located at $\pm \Omega_i$ around the carrier and correspond to transitions between the dressed states. The observation of transitions from the ground state to the lowest pair of dressed states, giving rise to the celebrated vacuum Rabi splitting, and the simultaneous observation of direct transitions between dressed states is enabled by two features of our system: First, although we achieve cooling close to the motional ground state~\cite{Meng18}, there is a finite population of the first excited states and, therefore, we can observe transitions starting from these states; second, the energy gap, $\hbar\Omega_i$, between the dressed states is comparable to the energy gap between the ground state and the first excited state manifold, $\hbar\omega_i$, so that the corresponding transitions have similar energy and can be detected by the same method. The position of these sidebands allows us to precisely measure the Rabi splitting, $\Omega_i$. We find $\Omega_y = 2\pi \times \SI{35(1)}{kHz}$  and $\Omega_x = 2\pi \times \SI{36(1)}{kHz}$. This corresponds to coupling strengths of $g_y/\omega_y = 0.19(1)$ and $g_x/\omega_x = 0.12(1)$, respectively. Thus, we clearly reach the ultra-strong coupling regime for both DOFs~\footnote{Here, we choose the $g_i/\omega_i$ ratio as a figure of merit for USC. Another common choice consists in using the $\Omega_i/\omega_i$ ratio~\cite{Guenter09,Todorov10,Schwartz11,George16}, which would yield, in our case, $\Omega_y/\omega_y = 0.38(1)$ and $\Omega_x / \omega_x = 0.24(1)$.}.

Another feature of our setup is the possibility to tune the coupling strength \textit{in situ}. For this purpose, we use an additional fiber-guided light field at the so-called tune-out wavelength, near \SI{880}{\nano\meter}~\cite{Arora11}. At this wavelength, the scalar polarizability vanishes, so that this  laser field only induces a vector light shift. This field propagates in the same direction as the blue-detuned trapping light field and has the same polarization. For this configuration, we expect a partial compensation of the fictitious magnetic field gradient~\cite{Albrecht16} and, thus, a reduction of the coupling strength. To experimentally quantify this effect, we measured the Rabi splitting, $\Omega_y$, for different powers of the tune-out laser, $P_{880}$, see Fig.~\ref{fig:880_effect}. As expected in our regime, $\Omega_y$ decreases linearly with $P_{880}$. The measured slope is $\mathrm{d}\Omega_y/\mathrm{d}P_{880} = \SI{-120(10)}{\hertz\per\micro\watt}$. From an \textit{ab initio} calculation, taking into account the vector polarizability of cesium and the mode function of the nanofiber-guided tune-out light, we expect $\SI{-100}{\hertz\per\micro\watt}$, in reasonable agreement with the experimental value. We note that, for a different experimental configuration, the tune-out laser field may also enhance the coupling strength. In this case, a power of $P_{880} \approx \SI{800}{\micro\watt}$ should be sufficient to induce a coupling on the order of the trap frequency.  Furthermore, by modulating the tune-out laser field intensity, one may dynamically adjust the coupling strength, even on timescales shorter than the Rabi oscillation period. This might enable, e.g., adiabatic USC/DSC ground-state preparation or the study of quench dynamics.

In summary, the demonstrated implementation of a mechanical analogue of the Dicke model with cold atoms constitutes a novel route to explore ultra-strong and, potentially, even deep-strong coupling phenomena with  unprecedented level of control. 
While our implementation takes advantage of the specific polarization gradients intrinsically present in our nanofiber-based optical trap, other optical micropotentials, such as optical lattices~\cite{Bloch05}, also qualify for implementing our scheme~\cite{Schneeweiss18}. Possible future research directions include the study of the dynamical Casimir effect via a modulation of the system parameters~\cite{Liberato07} or of the role of dissipation in the USC/DSC regime~\cite{Beaudoin11}. Understanding these effects will be beneficial, e.g., for the realization of ultra-fast quantum gates~\cite{Romero12,Kyaw15} or of qubit protection protocols~\cite{Nataf11} relying on USC. Finally, a suitably tailored real and fictitious magnetic field pattern can be used to realize generalizations of the quantum Rabi model or of the Dicke model, such as the driven QRM, or to implement ultra-strong two-photon coupling~\cite{Schneeweiss18}.

\section*{Acknowledgments}

Financial support by the European Research Council (CoG NanoQuaNt) and the Austrian Science Fund (FWF, SFB NextLite Project No. F 4908-N23 and DK CoQuS project No. W 1210-N16) is gratefully acknowledged.

\bibliography{bibliography}

\end{document}



\title{Supplementary Material: Cold-atom-based mechanical analogue of the Dicke model\\in the ultra-strong coupling regime}

\author{A.~Dareau}
\author{Y.~Meng}
\author{P.~Schneeweiss}
\email{philipp.schneeweiss@tuwien.ac.at}
\affiliation{%
 Vienna Center for Quantum Science and Technology,\\
 TU Wien -- Atominstitut, Stadionallee 2, 1020 Vienna, Austria
}%
\author{A.~Rauschenbeutel}
\email{arno.rauschenbeutel@hu-berlin.de}
\affiliation{%
 Vienna Center for Quantum Science and Technology,\\
 TU Wien -- Atominstitut, Stadionallee 2, 1020 Vienna, Austria
}%
\affiliation{%
 Department of Physics, Humboldt-Universit\"at zu Berlin, 10099 Berlin, Germany
}%

\date{\today}
\maketitle

\section{Experimental setup and sequence}

Laser-cooled cesium atoms are trapped in the evanescent field surrounding a silica optical nanofiber of nominal radius $a = \SI{250}{nm}$. The trapping potential is created by sending a blue-detuned running wave field with a free-space wavelength of \SI{783}{\nano\meter} and a power of $\SI{17.8}{\milli\watt}$ and a red-detuned standing wave field at \SI{1064}{\nano\meter} wavelength with a total power of $\SI{2.80}{\milli\watt}$ into the nanofiber. The blue- and the red-detuned fields are guided in the quasi-linearly polarized fundamental HE$_{11}$ modes and the  polarizations of the two fields are orthogonal. Two diametric arrays of trapping sites are formed~\cite{Vetsch10}. The calculated trap frequencies are $(\omega_x, \,\omega_y, \,\omega_z) = 2\pi\times(145,\,95,\,228)\,\mathrm{kHz}$ in the radial, azimuthal, and axial directions, respectively, for a total trap depth of about $\SI{200}{\micro\kelvin}$. The trap minima are located about $\SI{270}{\nano\meter}$ away from the nanofiber surface.

The atoms are loaded into the nanofiber-based trap from a magneto-optical trap via an optical molasses stage~\cite{Vetsch10}. In this process, the collisional blockade effect~\cite{Schlosser02} limits the maximum number of atoms per trapping site to one. After loading, atoms are distributed over the two diametric arrays of trapping sites. In order to reduce possible inhomogeneities, we remove atoms from one of the two arrays. This is achieved by performing degenerate Raman cooling using a fiber-guided light field, as first demonstrated in~\cite{Meng18}. The fiber-guided light at the position of the atoms is almost perfectly $\sigma^-$- and $\sigma^+$-polarized for one array and the other, respectively. This leads to degenerate Raman cooling of the atoms in one array and to heating of the atoms in the other array. After a few millisecond of optical pumping, all the atoms on the $\sigma^+$-side are lost, while the atoms on the $\sigma^-$-side remain trapped and are cooled close to the ground state.

The heterodyne fluorescence spectroscopy setup is described in~\cite{Meng18}: The excitation laser beam features a waist of about $\SI{1.2}{\milli\meter}$ at the position of the atoms and a total power of $\SI{120}{\micro\watt}$. Its frequency is red-detuned by about $\SI{55}{MHz}$ with respect to the $F=4 \to F^\prime=5$ transition of the $D_2$ line of cesium. The corresponding peak intensity is around $5\,\mathrm{I_\mathrm{sat}}$, where  $I_\mathrm{sat}$ is the saturation intensity of the considered transition. The resulting saturation parameter is on the order of $10^{-2}$, meaning that incoming photons are mostly coherently scattered. Atoms are exposed to the excitation beam for $\SI{100}{\milli\second}$. A typical spectrum as shown in Fig.~2 is obtained after averaging over about 7000 experimental runs.

For an offset magnetic field smaller than \SI{0.2}{G}, degenerate Raman cooling is no longer effective as the Zeeman states become degenerate in energy. The atoms are then heated out of the trap by the excitation laser in about \SI{10}{\milli\second} during heterodyne detection due to recoil heating, which lowers the signal to noise ratio. To mitigate this problem, we use an interleaved detection scheme alternating \SI{5}{\milli\second} of heterodyne measurement and \SI{2}{\milli\second} of optical molasses cooling. We perform between 20 to 40 cycles for each experimental run. This interleaved detection scheme was used for the spectra corresponding to Zeeman splittings of $\SI{0}{\kilo\hertz} \leq \Delta/(2\pi) \leq \SI{70}{\kilo\hertz}$ in Fig.~2c, i.e., for $\SI{0}{G} \leq B_0 \leq \SI{0.2}{G}$. We compared the spectra obtained with this interleaved scheme with the ones measured with the continuous scheme for ranges of offset magnetic field where both methods work. We did not observe a significant difference on the peak positions for the two schemes.

\section{Detailed derivation of the spin-motion coupling}

Here, we detail the mapping of Hamiltonian~(1) to Hamiltonian~(2). Assuming a linear gradient for the fictitious magnetic field ($\bm{B}_\mathrm{fict} \approx b_y y \,\bm{e}_x$) and using that $\bm{B}_0 = B_0\,\bm{e}_y$, Hamiltonian~(1) becomes:
\begin{equation}
 \hat{H}_y =  \hbar \omega_y \adyop \ayop + g_F \mu_B B_0 \Fy + g_F \mu_B b_y \hat{y}\Fx.
\label{eq:Hat_QRM_spin_position}
\end{equation}

\noindent We can write the position operator $\hat{y}$ in terms of the raising and lowering operators using $\hat{y} = y_0 (\adyop + \ayop)$, where $y_0 = \sqrt{\hbar/(2M\omega_y)}$ is the size of the harmonic oscillator ground state. Here $M$ denotes the atomic mass. We also introduce $\Fp$ and $\Fm$, which are the raising and lowering operators for the eigenstates of $\Fy$, respectively, so that $\Fx = (\Fp + \Fm) / 2$. Hamiltonian~(\ref{eq:Hat_QRM_spin_position}) can then be written as:
\begin{equation}
 \hat{H}_y =  \hbar \omega_y \adyop \ayop + g_F \mu_B B_0 \Fy + \frac{g_F \mu_B b_y y_0}{2} \left( \adyop + \ayop \right) \left( \Fp + \Fm \right).
\end{equation}

\noindent This is identical to Hamiltonian~(2), where we identify the Zeeman splitting, $\Delta$, and the spin-motion coupling strength, $g_y$, as:
\begin{align}
 \hbar\Delta & = g_F \mu_B B_0, \\
 \frac{\hbar g_y}{\sqrt{2}} & = \frac{g_F \mu_B b_y y_0}{2}.
\end{align}

We denote $\ket{m_F, n_y}$ the eigenstates of Hamiltonian~(2) in the absence of spin-motion coupling, where $m_F$ and $n_y$ label eigenstates of $\Fy$ and Fock states of the harmonic potential along $y$, respectively. We are interested in the Rabi frequency, $\Omega_y$, between the low-energy states $\ket{a} = \ket{m_F=-F, n=1}$ and $\ket{b} = \ket{m_F=-F+1, n=0}$. For $\Delta = \omega_y$, these states are resonantly coupled and we find:

\begin{equation}
\frac{\hbar\Omega_y}{2} = \left|\bra{b} \hat{H}_y \ket{a} \right| = \frac{\hbar g_y}{\sqrt{2F}} \left|\bra{b} \ayop \Fp \ket{a} \right| = \hbar g_y.
\end{equation}

\noindent Hence, we find $\Omega_y = 2 g_y$.

\section{Models for numerical simulations} 
\subsection{Simplified model}

Here, we describe the derivation of the expected transitions frequencies, indicated by black lines in Fig.~3c. In order to gain more intuition, we used a simplified model, only considering the low-energy states shown in Fig.~3b. This model turns out to be sufficient to describe the observed spectra. Indeed, because of the degenerate Raman cooling and optical pumping originating from the probing laser field, most of the atoms populate these low-lying energy states. The simplified model only considers the following bare states:
\begin{align}
 \ket{g} & = \ket{m_F=-4, n_x=0, n_y=0}, \\
 \ket{e} & = \ket{m_F=-3, n_x=0, n_y=0}, \\
 \ket{x} & = \ket{m_F=-4, n_x=1, n_y=0}, \\
 \ket{y} & = \ket{m_F=-4, n_x=0, n_y=1}.
\end{align}

\noindent The corresponding Hamiltonian reads:
\begin{equation}
 \begin{split}
 \hat{H}/\hbar = & \Delta \ketbra{e}{e} + \omega_x \ketbra{x}{x} + \omega_y \ketbra{y}{y} \\
        + & (\Omega_x/2) \left(\ketbra{e}{x} + \ketbra{x}{e} \right) \\
        + &(\Omega_y/2) \left(\ketbra{e}{y} + \ketbra{y}{e} \right),\\
 \end{split}
\label{eq:simple_Ham}
\end{equation}

\noindent where $\Omega_{x,y} = 2 g_{x,y}$ are the Rabi frequencies for the $x\leftrightarrow e$ and $y\leftrightarrow e$ transitions, respectively. A numerical diagonalization of~(\ref{eq:simple_Ham}) in the basis $\{ \ket{g}, \ket{e}, \ket{x}, \ket{y} \}$ for different values of the Zeeman splitting, $\Delta$, yields sets of four eigenergies. The fluorescence spectra, as presented in Fig.~2, are expected to show peaks at frequencies corresponding to the difference of the obtained eigenenergies. In order to confirm the validity of this simplified model, we compared its results with the ones of a numerical simulation considering the full Hamiltonian, see supplementary material. In our range of parameters and for the typical temperatures measured in our setup, the two models show a good agreement.

The fit of the experimental data with this model is performed as follows: Trap frequencies are inferred by fitting the peaks corresponding to motional sidebands for large values of the offset magnetic field, $B_0$, where the effect of spin-motion coupling is negligible. More in detail, we perform a fit of the spectra shown in Fig.~2c for $\SI{250}{\kilo\hertz} \leq \Delta/(2\pi) \leq \SI{330}{\kilo\hertz}$. The Zeeman splitting, $\Delta$, was calibrated as follows: The coils generating the offset magnetic field, $B_0$, were calibrated independently using microwave spectroscopy. From this calibration, we inferred the linear relation between the current circulating in the coils and the amplitude of the resulting magnetic field. In order to account for shifts induced by spurious real and fictitious magnetic fields originating, e.g., from the excitation light field, we performed a fit on the peak corresponding to the $\ket{g} \to \ket{e}$ transition in Fig.~2c. Its position corresponds to $\omega_S-\omega_I = \Delta$. Again, this was done for large values of $B_0$, corresponding to $\Delta / (2\pi) \geq \SI{270}{\kilo\hertz}$. Finally, the coupling strengths $g_{x,y}$ were obtained by fitting the position of the peak corresponding to the transition between dressed states at resonance, $\Delta = \omega_{x,y}$, as explained in the main text of the manuscript.

\subsection{Full Dicke model} 

\begin{figure}
	\includegraphics[width=\columnwidth]{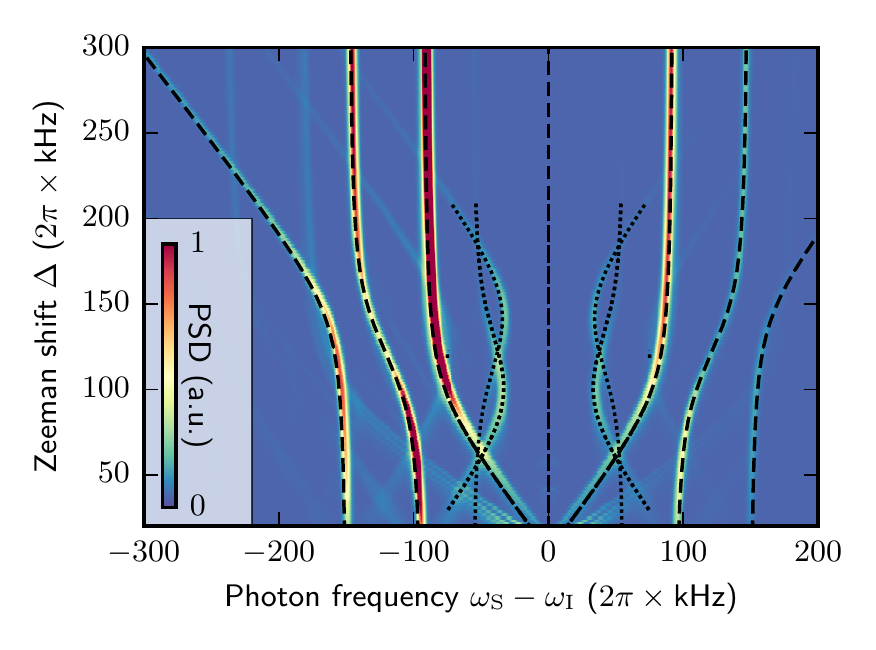}
	\caption{\footnotesize \textbf{Comparison between simplified and full theoretical model}. The black dashed lines correspond to the simplified model discussed above, only considering low-energy states. They are identical to the ones shown in Fig.~2c. The result of a numerical diagonalization of the full Hamiltonian is shown with a colormap for the parameters given in Table.~\ref{table:simulation_parameters}. The two models agree very well.}
	\label{fig:Methods_simulation_comparison}
\end{figure}

\begin{table}
\footnotesize
  \vspace{0.5cm}
  {\renewcommand{\arraystretch}{1.2}
  \begin{tabular}{|l|c|c|}
  \hline
    \textbf{Parameter} & \textbf{Notation} & \textbf{Value}\\
    \hline
    \multicolumn{3}{|c|}{$y$-DOF} \\
    \hline
    trap frequency & $\omega_y$ & $2\pi \times \SI{93}{kHz}$ \\
    coupling strength & $g_y$ & $2\pi \times \SI{17.5}{kHz}$ \\
    mean number of motional quanta  & $\mean{n_y}$ & $0.5$ \\
    Lamb-Dicke parameter  & $\eta_y$ & $0.15$ \\
    \hline
    \multicolumn{3}{|c|}{$x$-DOF} \\
    \hline
    trap frequency & $\omega_x$ & $2\pi \times \SI{149}{kHz}$ \\
    coupling strength & $g_x$ & $2\pi \times \SI{18}{kHz}$ \\
    mean number of motional quanta  & $\mean{n_x}$ & $0.5$ \\
    Lamb-Dicke parameter  & $\eta_x$ & $0.1$ \\
    \hline
  \end{tabular}
  }
  \caption{\footnotesize Parameters used for the numerical simulations shown in Fig.~\ref{fig:Methods_simulation_comparison} and inferred from the experimental spectra shown in Fig.~2c.} 
\label{table:simulation_parameters}
\end{table}

Here, we compare the simplified model, described in the methods and used to compute the expected transitions frequencies in Fig.~2c, with a full diagonalization of Hamiltonian~(2), taking into account the $x$ and $y$ motional DOFs as well as all the Zeeman states. The result is shown in Fig.~\ref{fig:Methods_simulation_comparison}. We performed a numerical diagonalization of~(2) using the QuTiP python toolbox~\cite{Johansson13}, for the parameters shown in Table~\ref{table:simulation_parameters}. The simulation was considering a $F=4$ spin, and a set of $5$ motional states for each motional DOF. We assumed a temperature corresponding to a mean number of motional quanta of $\mean{n} = 0.5$~\cite{Meng18}. From the numerical diagonalization, we obtain a set of eigenenergies $\{E_i\}$ and corresponding eigenstates $\{\ket{i}\}$. The transition $\ket{i}\to \ket{j}$ gives rise to a peak centered at a frequency $\hbar\omega_{ij} = E_i - E_j$. For the spectra plotted in Fig.~\ref{fig:Methods_simulation_comparison}, we assumed the peaks to be Gaussian with a constant width of \SI{2}{kHz}. The amplitude of the peak is given by the initial (thermal) population of the state $\ket{i}$ and a Frank-Condon factor $F_{ij} = \bra{j}\hat{V}\ket{i}$. In the Lamb-Dicke regime, we can write $\hat{V} = (1 + \eta_x [\adxop + \axop] + \eta_y [\adyop + \ayop]) \otimes \hat{S}_- \hat{S}_-^\dagger$, where $\eta_{x}$ and $\eta_{y}$ are the corresponding Lamb-Dicke parameters. Here, $\hat{S}_-$ ($\hat{S}_-^\dagger$) is a generalized lowering (raising) operator for the emission (absorption) of a $\sigma^-$-polarized photon, taking into account the Clebsch-Gordan coefficients of the considered optical transition~\cite{Gangl02}. As shown in Fig.~\ref{fig:Methods_simulation_comparison}, the agreement between the simplified model (dashed lines) and the full model (colormap) is very good for the range of parameters considered here.

\bibliography{bibliography}